\documentclass{mem}
\usepackage{natbib}
\usepackage{txfonts}
\usepackage{balance}
\usepackage{graphicx}
\usepackage{txfonts}
\usepackage[a4paper]{hyperref}
\idline{79}{3}

\newcommand{\ie}{\textit{i.e.},}
\newcommand{\eg}{\textit{e.g.},}
\newcommand{\Ms}{\ensuremath{M_\odot}}

\newcommand{\el}[2]{$\rm{}^{#2}\kern-0.6pt#1$}

\usepackage{color}

\begin{document}
\def\teff{$T\rm_{eff }$}
\def\kms{$\mathrm {km s}^{-1}$}

\title{Massive stars, globular clusters and elliptical galaxies
}

   \subtitle{}

\author{
G. \,Meynet\inst{1} 
T. \, Decressin\inst{1}
C. \, Charbonnel\inst{1,2}          }

  \offprints{G. Meynet}

\institute{
Geneva Observatory --
University of Geneva,
CH-1290 Sauverny, Switzerland
\and
Laboratoire d'Astrophysique de Toulouse et Tarbes, 
CNRS UMR 5572, OMP, 
Université Paul Sabatier 3, 14 Av. E. Belin, 
31400 Toulouse, France
}

\authorrunning{Meynet }

\titlerunning{Massive rotating stars}

\abstract{Globular clusters as $\omega$ Cen and NGC 2808 appear to have a population of very He-rich stars.
From a theoretical point of view, one expects the presence of He-rich stars in all globular clusters showing an oxygen-sodium (O-Na) anticorrelation. 
In this paper, we briefly recall how fast rotating massive stars could be the main source of the material from which He-rich stars have formed. 
We speculate that the UV-upturn phenomenon observed in all elliptical galaxies might be due to He-rich stars.
If this hypothesis is correct then
fast rotating massive stars may have played in the early evolutionary phases of these systems a similar role as the one they played in the nascent globular clusters.
}
\maketitle{}

\section{Chemical anomalies in globular clusters}

The finding of a double sequence in the globular cluster $\omega$ Cen by \citet{Anderson97} 
and the  further interpretation of the bluer sequence by a
strong excess of helium  constitutes a major enigma  for stellar and
galactic evolution \citep{Norris04}. 
The bluer sequence with a metallicity
[Fe/H]= -1.2  or $Z =  2 \cdot 10^{-3}$ implies an He--content $Y=0.38$ (0.35-0.45), 
i.e. an helium enrichment $\Delta Y = 0.14$ \citep[see also][]{Norris04}.
In turn, this demands a relative helium 
to metal enrichment  $\Delta Y/\Delta Z$ of the order of   70
\citep{Piotto05}. 
This is enormous and more than one order of magnitude larger than the current value of $\Delta Y/\Delta Z= 4-5$ \citep{Pagel92} obtained from  
  extragalactic  HII regions.  A value of 4-- 5  is  consistent with the chemical yields from supernovae
\citep{Maeder92} forming  black holes above about 20--25 M$_{\odot}$. \cite{MMocen} showed that the wind contributions of low $Z$ rotating stars  are able to produce the high $\Delta Y/\Delta Z$
 observed in the  blue Main Sequence in  $\omega$ Cen.

Recent observations show that He-rich stars in globular clusters might be a common feature.
For instance, clusters like NGC 2808, M13 and NGC 6441 \citep{Caloi07} show a well-developed blue horizontal branch, and a strong slope upward from the red clump to the blue of the RR Lyrae regions. Both features could be explained if He-rich stars are present in these clusters.  
\citet{Kaviraj07} studied the UV and optical properties of 38 massive globular clusters in M87
(elliptical galaxy in virgo cluster). A majority of these clusters appear extremely bright in the far-UV. Canonical models (i.e. models without any super He-rich stars) would imply ages for these clusters about 3 - 5 Gyr larger than the age of the Universe according to the WMAP data! This difficulty is removed when a  super-He-rich ($\Delta Y/\Delta Z > \sim 90$) stellar component is supposed to be present in these clusters.

It has also long been known that globular cluster stars present some
striking anomalies in their content in other light elements\footnote{On the contrary, the content in heavy
elements (i.e., Fe-group, $\alpha$-elements) is fairly 
constant from star to star in any well-studied 
individual Galactic globular cluster (with the notable exception of $\omega$~Cen).}: while in all the
Galactic globular clusters studied so far one finds ``normal" stars with detailed chemical 
composition similar to that of field stars of same metallicity (i.e., same [Fe/H]), 
one also observes numerous ``anomalous" main sequence and red giant stars that are 
simultaneously deficient (to various degrees) in C, O, and Mg, and enriched in N, Na, and Al 
(for recent reviews see \citealp{GrattonSneden2004,Charbonnel2005}). 
Additionally, the abundance of the fragile Li was found to be anticorrelated with
that of Na in turnoff stars in a couple of globular clusters 
(\citealp{PasquiniBonifacio2005,BonifacioPasquini2007}).

It is clear now that these chemical peculiarities have been inherited at birth 
by the low-mass stars we observe today in globular clusters, and that their root cause 
is H-burning through the CNO-cycle and the NeNa- and MgAl-chains 
that occurred in an early generation of more massive and faster evolving 
globular cluster stars.
In other words, compelling evidence leads us to believe that at least two generations 
of stars succeeded in all Galactic globular clusters during their infancy. The first one 
corresponds to the bulk of ``normal'' stars born with the pristine composition 
of the protocluster gas; these objects are those with the highest O and Mg and 
the lowest He and Na and Al abundances also found in their field contemporaries. 
The second generation contains the stars born out of material
polluted to various degrees by the ejecta of more massive stars, and which present
lower O and Mg and higher He, Na and Al abundances than their first generation counterparts.

Any model aiming at explaining the chemical 
properties of globular cluster stars should give an answer to the following questions: 
(1) Which type of stars did produce the material enriched in H-burning products? 
(2) What is the physical mechanism responsible for selecting only material bearing the 
signatures of H-processing?
(3) Why does this process occur only in globular clusters?\footnote{Indeed up to now
the peculiar chemical patterns observed in globular clusters, i.e., the O-Na and Mg-Al 
anticorrelations, have not been found in field stars.}
\citet{PC06} and
\citet{DecressinI, DecressinII} propose answers to the above three questions. The main line of the reasoning is recalled below. 

\section{Slow winds of fast rotating massive stars}

Fast rotating massive stars, with an initial metallicity $Z=0.0005$, with typical time averaged velocity of 500~km~s$^{-1}$ on 
the main sequence easily reach the critical velocity\footnote{The critical velocity is the velocity at the equator 
such that the centrifugal acceleration balances the gravity.} 
at the beginning of their evolution and remain near the critical limit during the rest
of the main sequence and part of the central He-burning phase
\citep{DecressinI,Ekstrom07}. 
As a consequence, during those phases they lose large amounts of material 
through a mechanical wind\footnote{For example, our 60~M$_{\odot}$ model 
loses 24.3~M$_{\odot}$ in its slow wind.}, which probably leads to the formation of a slow 
outflowing keplerian equatorial disk. 
This is the kind of process believed to occur in Be stars \citep{PorterRivinius2003}. 
 
The material ejected in the disk has two interesting characteristics: (1) due to
rotational mixing that transports the products of the CNO cycle and of the
Ne-Na and Mg-Al chains from the core to the stellar surface, it bears the signatures  
of H-burning and presents abundance patterns similar to the chemical anomalies 
observed in the second generation stars \citep[see for more details][]{DecressinI}\footnote{We remind however that in order to reproduce
the lowest Mg values observed we had to postulate an increase of the $^{24}$Mg(p,$\gamma$)$^{25}$Al nuclear
reaction rate with respect to the published values.};
(2) it is released into the circumstellar environment
with a very low velocity and thus can easily be retained into the shallow potential well 
of the globular cluster. In the following we shall call it the {\it slow wind}. 

\begin{figure}[t]
\resizebox{\hsize}{!}{\includegraphics[clip=true]{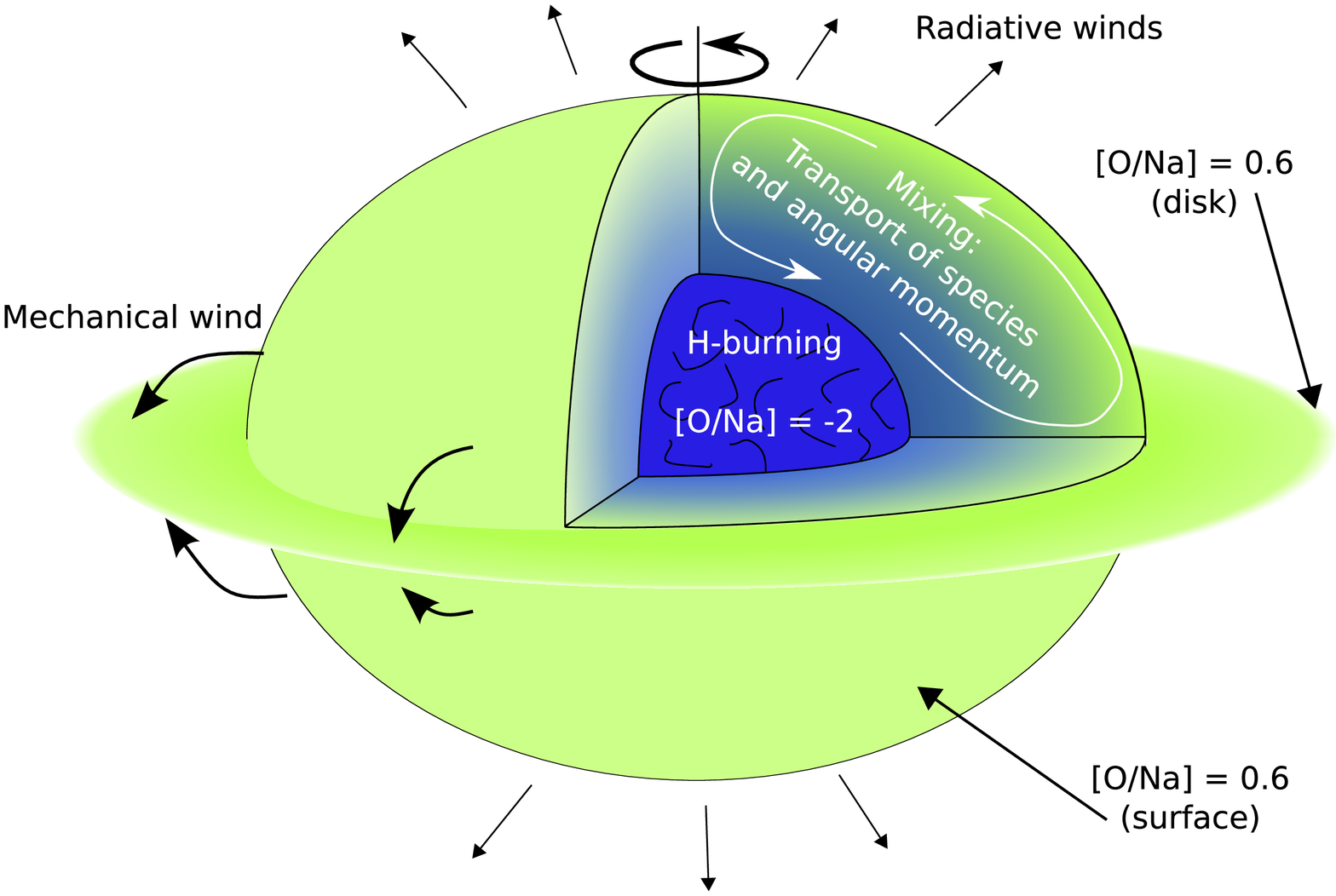}}
\resizebox{\hsize}{!}{\includegraphics[clip=true]{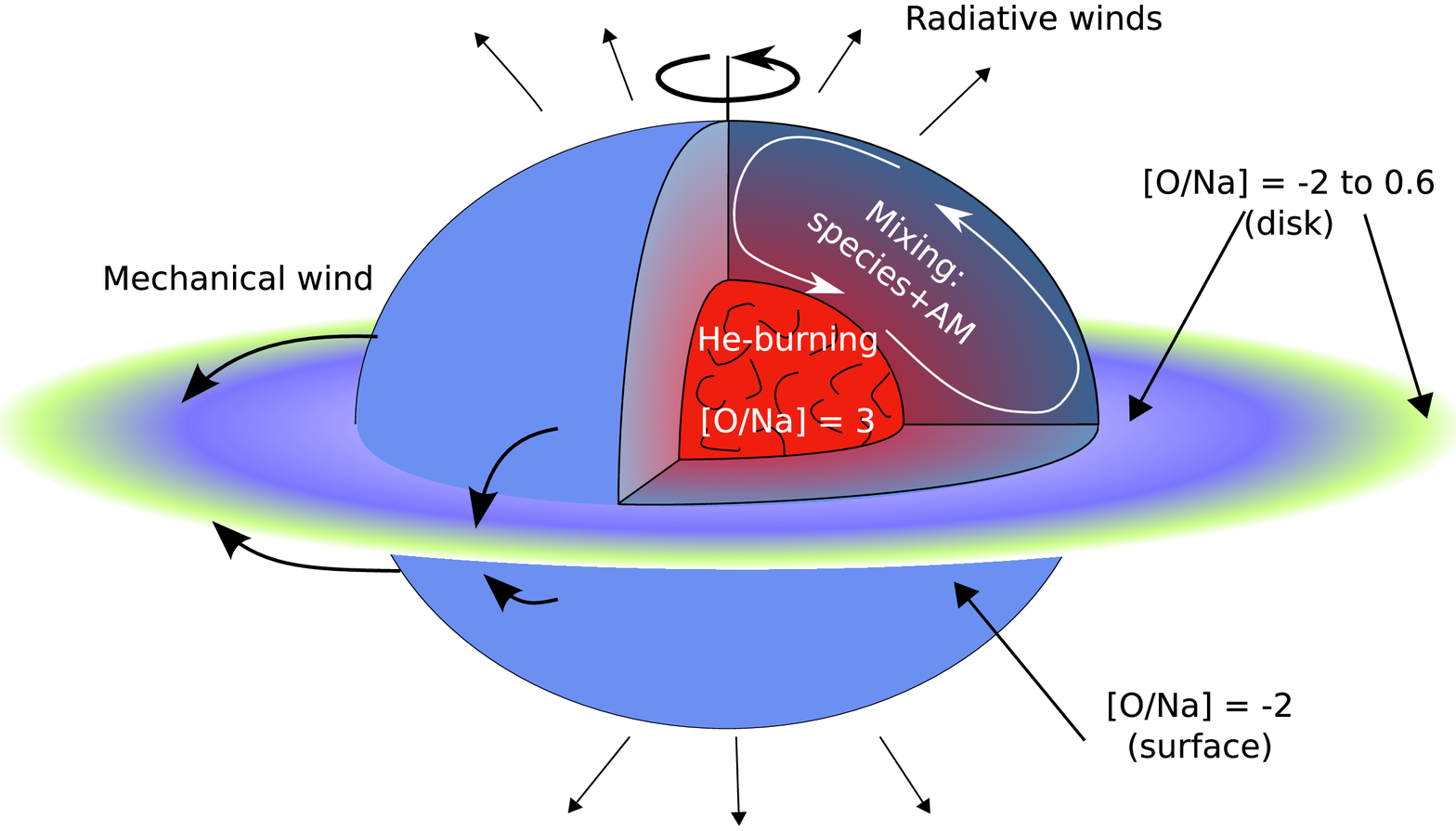}}
\resizebox{\hsize}{!}{\includegraphics[clip=true]{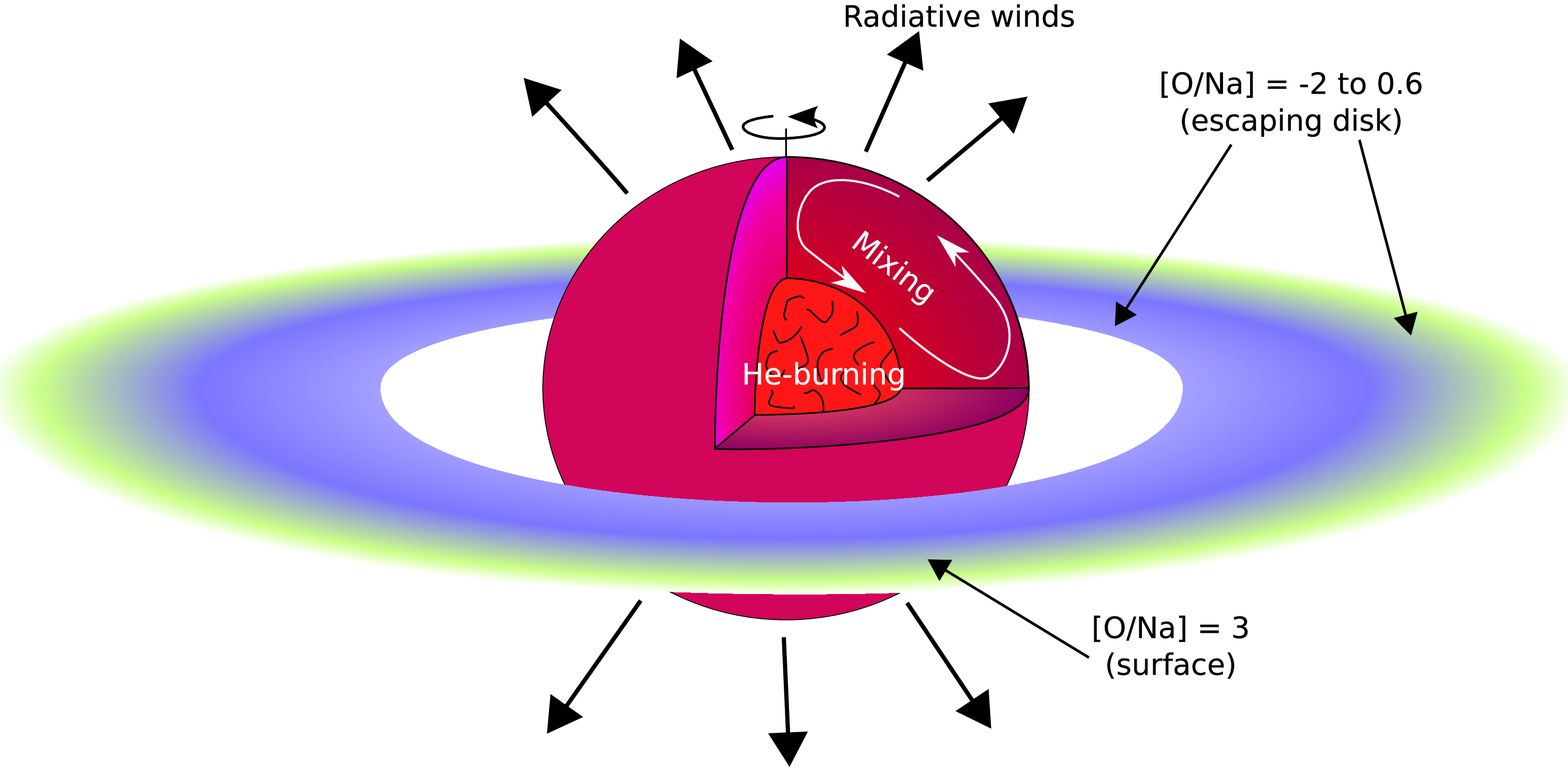}}
\caption{\footnotesize Schematic view of the evolution of fast rotating massive
    stars. The colours reflect the chemical composition of the various 
    stellar regions and of the disk. 
    (top) During the main sequence, a slow outflowing equatorial disk forms 
    and dominates matter ejection with respect to radiative winds.
    Early on 
the main sequence, the [O/Na] value in the slow wind is the pristine one. 
    (middle) As mixing proceeds, 
matter with [O/Na] typical of H-burning (i.e., $\sim$~-2) is ejected in the slow wind
    (bottom) Due to heavy mass loss, the star moves away from critical velocity
    and does not supply its disk anymore; radiatively-driven fast wind takes over 
    before the products of He-burning reach the stellar surface. Figure taken from
    \citet{DecressinII}.
}
\label{schemes}
\end{figure}

The disk-star configuration lasts during the whole main sequence for stars close 
to or at critical velocity; for the most massive stars that are at the $\Omega\Gamma$-limit (\ie{} when the
surface luminosity is near the Eddington limit), it persists on the LBV phase. 
When the star evolves away from the critical limit due to heavy mass loss, the radiatively-driven fast wind 
takes over and the disk is not supplied by the star anymore. This happens during the central He-burning phase, 
before the He-burning products reach the stellar surface and contaminate the slow wind component 
(bottom panel of Fig.~\ref{schemes}).  From that moment on the high-speed material ejected by the star 
and by the potential supernova will escape the globular cluster.
\textit{We proposed that this filtering mechanism, which retains in the potential well 
  of the globular cluster only the slow stellar ejecta, is the physical mechanism responsible 
  for supplying the required H-processed material for forming the stars of second generation.} 

The large star-to-star spread in light
elements observed today indicates that all the material ejected by the
first generation massive stars did not have the time to be fully mixed
before being recycled in the second generation. In the frame of our model,
it imposes
(see also Sects.~3 and 4.3), that 
star formation occurred 
in the vicinity of individual massive rotating star, from clumps 
made of slow wind material and of
pristine interstellar gas. This star formation may be triggered for
instance by the ionisation front. 

\begin{figure}[t]
\resizebox{\hsize}{!}{\includegraphics[clip=true]{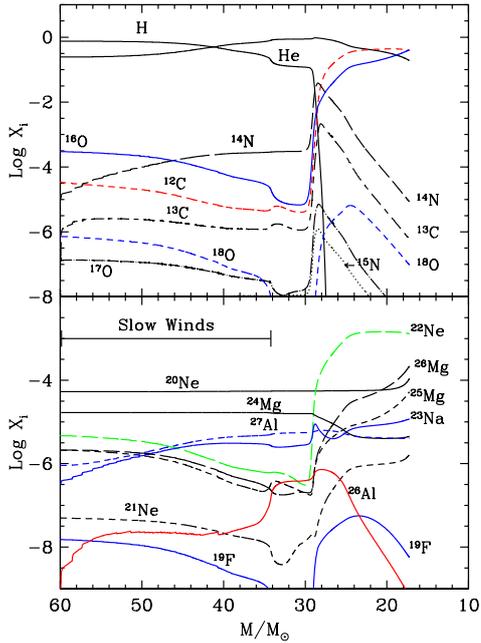}}
\caption{\footnotesize Evolution of the surface abundances in mass fraction as a function of the remaining
mass for a 60 M$_\odot$ stellar model at $Z=0.0005$ and an initial rotational velocity on the ZAMS of 800 km s$^{-1}$. The  horizontal line labeled ``Slow Winds'' in the lower panel 
extends over the part of the evolution during which the rotating star releases slow equatorial winds. Second
generation stars are formed from clumps made of this material and of pristine interstellar medium.
We can see how as a function of time the abundance of oxygen in the slow wind decreases while that of
sodium increases.
}
\label{abon}
\end{figure}

The material ejected in the slow wind
presents various degrees of enrichment in H-burning products. At the
beginning of the disk-star phase, pristine material is ejected giving birth
to stars with a composition similar to that of the first generation. As
time proceeds, the slow wind becomes more and more polluted in H-burning
products (see Fig.~1 and \ref{abon}) and forms stars that are more and more ``anomalous''.
The slow wind phase lasts only a few million years
(\eg{} 4.5~Myr for a 60~\Ms{} star). 

\section{Constraints on dilution factor, IMF slope and degree of evaporation}

How many massive fast rotators are needed in order to reproduce the observed population of
anomalous stars in globular clusters? This question is first addressed by 
\citet{PC06} for NGC 2808 and 
in more details in
\citet{DecressinII} on the basis of theoretical stellar yields for NGC 6752. Here we just recall the method used as well as the main results for NGC 6752.

From rotating massive star models we can compute the amount of material ejected by slow winds with a given
chemical composition \citep[see for details][]{DecressinI}. 
As already discussed in \S~2, there is compelling evidence that
the slow wind of an individual star pollutes only on a small scale around
its stellar progenitor, and that it is diluted locally with pristine ISM.
In order to constrain quantitatively this local dilution process we use the
Li variation detected in NGC~6752 by \citet{PasquiniBonifacio2005}.
It is important to recall that the massive star ejecta are totally
``Li-free'', as this fragile element is destroyed in these objects.
We suppose that the matter ejected by an individual massive star early on
the main sequence encounters more pristine gas and is more diluted than the
matter ejected later. 
This produces a Li-Na anticorrelation. We chosed the dilution factor
in order to reproduce 
the Li-Na anticorrelation observed by \citet{PasquiniBonifacio2005}.
Doing so we obtain that
the total mass of second generation stars amounts to
about twice the total mass released under the form of slow winds by massive
stars.

Given a dilution factor, we can compute how many small mass stars 
can be formed from the slow wind of a given fast rotating massive star.
With an Initial Mass Function (IMF), we can estimate the total number of second generation low mass stars
as well as the number of normal first generation low mass stars. 
Thus, the number ratio of anomalous to normal stars {\it at the time of birth of the second generation stars} can be predicted. During its evolution, a globular cluster may lose stars, for instance by tidal stripping.
The ratio of anomalous to normal star in that case will change as a function of time. 
Therefore, the number ratio of anomalous to normal stars {\it in present day cluster} 
depends on three parameters:
the dilution factor, the slope of the IMF and the number of stars which have escaped the cluster (evaporation).

As mentioned above, the dilution factor has been calibrated in order to reproduce the
Li-Na anticorrelation observed by \citet{PasquiniBonifacio2005}. Its value only depends on the stellar models used.
To constrain the IMF and the degree of evaporation we have only one observable, 
the present day observed fraction of anomalous to
normal stars in globular clusters  \citep[this value is equal to about 6 in NGC 6752,][]{CarrettaBragaglia2007}.
Thus we need to make some additional hypotheses to constrain these two quantities.

In scenario I we assume that the cluster underwent no evaporation of stars. In that case the present day observed fraction of anomalous to
normal stars gives directly access to the fraction at the time when the second generation was born. In this scenario the IMF can be constrained and the degree of evaporation is set equal to zero.

In our Scenario II, we assume that the massive polluters of first generation 
were born in the center of the cluster or have migrated very rapidly toward this region.
Then the second generation stars are created only in the central region while
the external part of the cluster hosts only first generation long-lived low-mass stars. 
This strong radial distribution is assumed to stay in place until the moment where supernovae 
sweep away the residual intra-cluster gas and thus strongly modify the cluster potential well. 
Evaporation will then preferentially remove normal stars in the outer layers of the cluster. 
In the Scenario~II, we impose a slope of the IMF 
equal to 1.35 and the degree of evaporation of normal stars is
constrained (no anomalous stars are supposed to have escaped the cluster).

To produce a high number of polluted stars (typically 5.7
times the number of normal stars as observed in NGC 6752) we need either
(a) in Scenario~I, a flat IMF with a slope for massive stars around 0.55 or
(b) in Scenario~II, that about 96\% of the unpolluted normal low-mass stars has
been lost by the cluster.

Using the two sets of free parameters obtained in Scenarios I and II, we then
explored various consequences which are briefly recalled below \citep[see for more details][]{DecressinII}.

\begin{figure}[t]
\includegraphics[width=2.5in,height=2.5in]{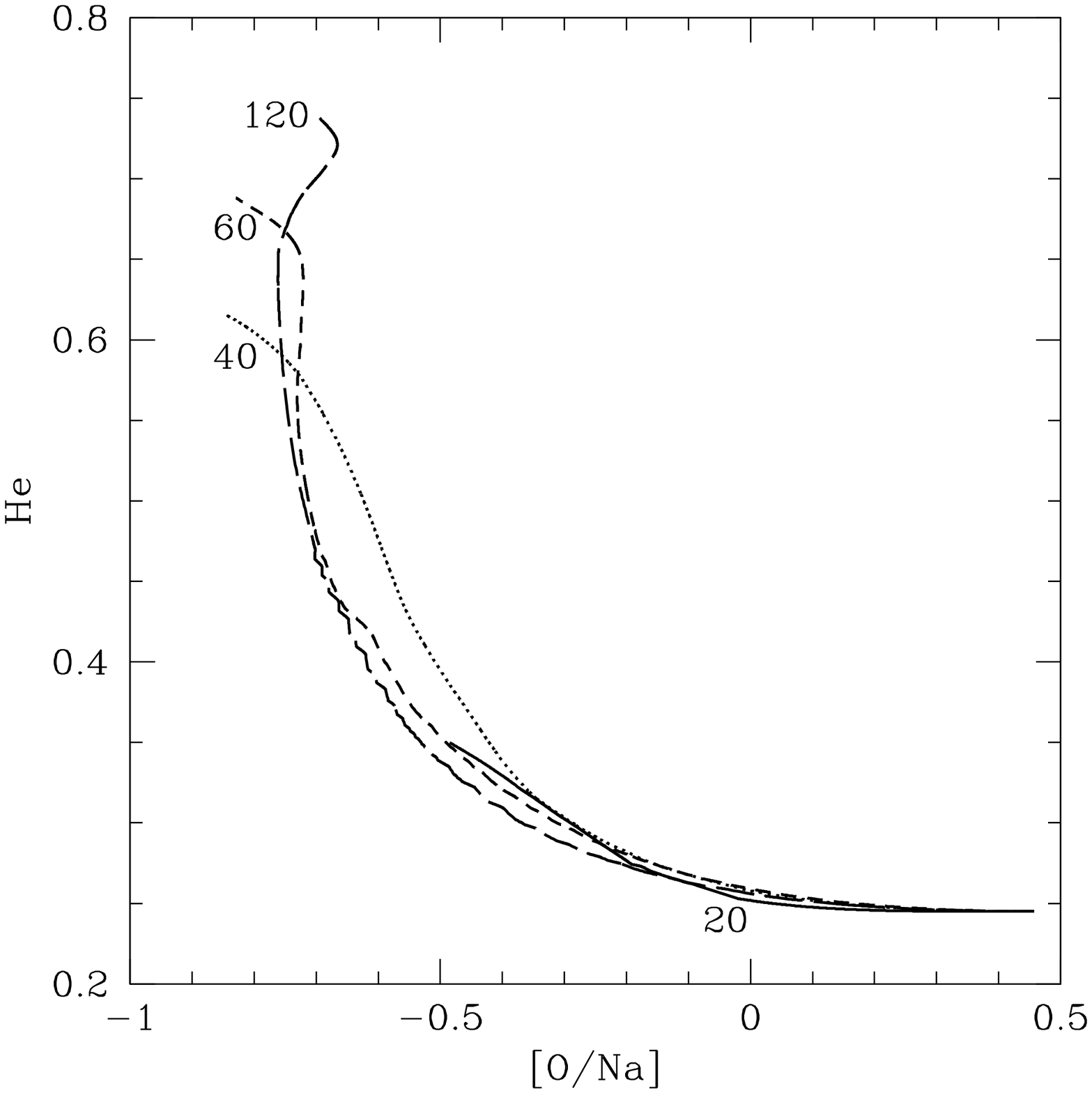}
\hfill
\includegraphics[width=2.5in,height=2.5in]{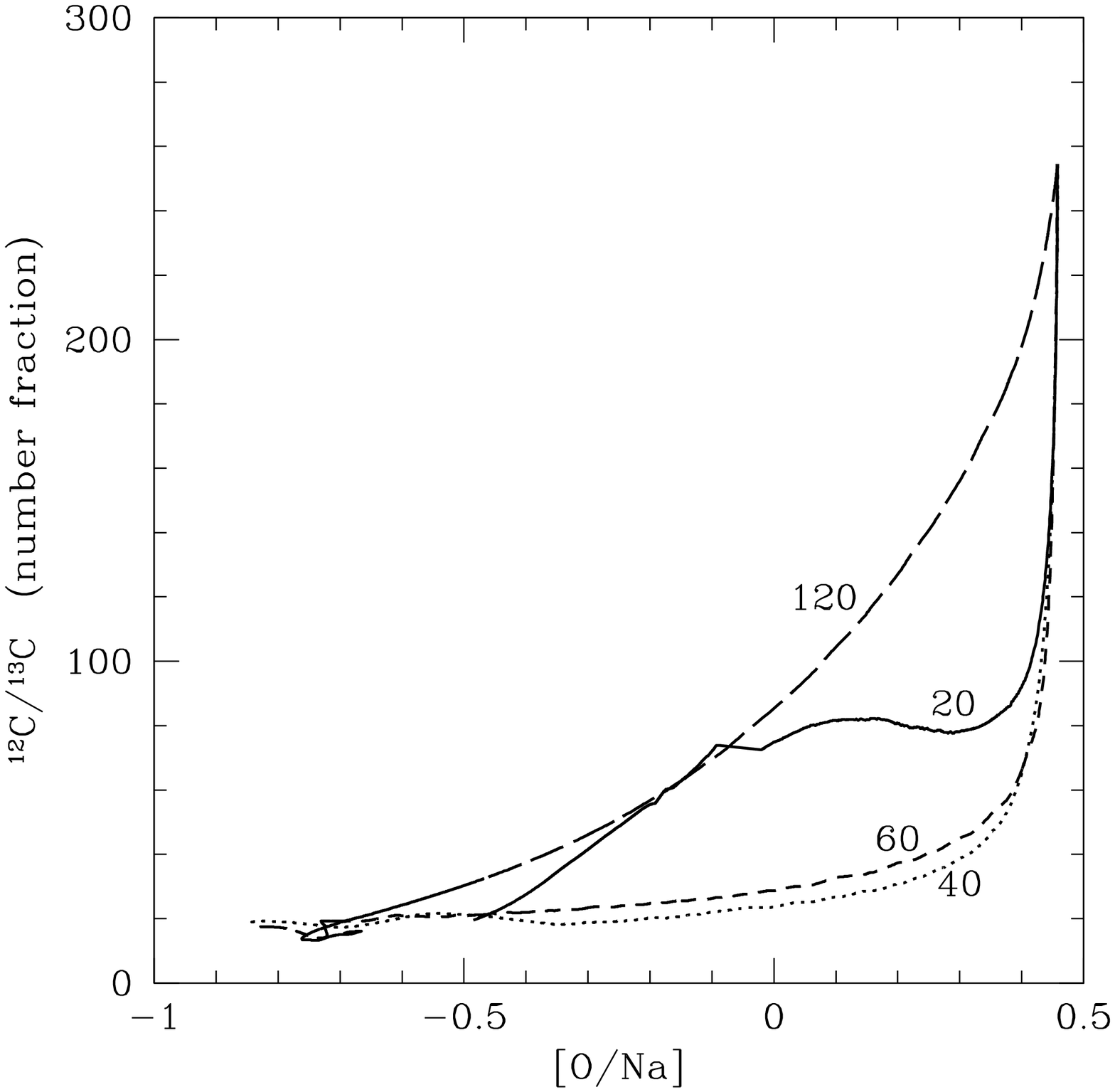}
\caption{\footnotesize {\it Upper panel}: Expected anticorrelation between helium and [O/Na] for
  second generation stars (at birth!) born from material ejected through slow winds of different
  initial mass models. 
{\it Lower panel}: Same as upper panel for the correlation between
  $^{12}$C/$^{13}$C and [O/Na]. Figure taken from
    \citet{DecressinII}}
\label{anticor}
\end{figure}

1) In both scenarios, we can produce an [O/Na] anticorrelation in the
  range $-1 \le {\rm [O/Na]} \le 0.62$. Quite remarkably, this range
  corresponds to the one observed in NGC 6752 if one considers only the set
  of stars for which both measurements of oxygen and sodium are
  available. Super-O-poor stars whose presence is indirectly
  inferred in some globular clusters need lower dilution coefficient or
  some additional mixing occurring in the star itself.
  
2) In both scenarios, around 12\% of
  the low-mass stars are expected to display a He content higher than 0.4
  (in mass fraction). Note that in the present model, He-enrichments are obtained keeping the sum
  C+N+O equal to its initial value as requested by the observations \citep{CohenMelendez2005, Carretta2006}.
  
3) An anticorrelation is predicted between the helium abundance and
  the value of the ratio [O/Na] at the surface of turnoff stars 
  (see upper panel of Fig.~\ref{anticor}). 
  
4) More than 30\% of stars are expected to have surface
  \el{C}{12}/\el{C}{13} ratios between 10 and 30.
  
5) The ratios $^{12}$C/$^{13}$C and [O/Na] are correlated (see lower panel of Fig.~\ref{anticor})\footnote{These correlations are obtained for non-evolved stars. When the star evolve along the red giant branch, internal mixing may change somewhat the results.}.  All the
  stars with [O/Na] ratios below about -0.6 are predicted to show values of
  the $^{12}$C/$^{13}$C ratios between 15 and 20.
  
6) Only a few percents of stars present at birth \el{O}{16}/\el{O}{17}
  ratios of the order of a few hundreds.
  
7) A correlation with a very large dispersion is predicted between the
  ratios \el{O}{16}/\el{O}{17} and [O/Na]. If only stars more massive
  than about 60~\Ms{} would contribute to the building up of the
  chemical anomalies in globular clusters, then only values above 2000
  would be expected would be expected.
  
8) A correlation between [C/N] and [O/Na] is expected, whose dispersion
  increases at lower [O/Na] values.
  
9) In the case of Scenario I, the present day mass cluster represents about half
  of the mass which has been used to form stars. The rest consists in
  matter ejected by stars and lost by the cluster.  The mass locked into
  remnants represents 19\% of the mass of the present day cluster, the
  first and second generation stars respectively 12 and 69\%.
  
10) In the case of Scenario II, the present day mass cluster represents only about 9\% of
  the mass which has been used to form stars. The rest consists in matter
  ejected by stars and of stars and stellar remnants lost by the cluster.
  The mass locked into remnants represents 8\% of the mass of the present
  day cluster, the first and second generation stars respectively 16 and
  76\%.

The above estimates are based on observed caracteristics of NGC 6752.
Using other clusters for calibrating the free parameters would of course
produce different results.
  
\section{Discussion}  

\subsection{Why no anomalous stars in the field of the halo?}

Globular clusters contain a significant population of anomalous stars.
In the field, at the moment, only normal stars are observed at the corresponding metallicity
(no anticorrelation found). 
This indicates that something special occurs in clusters which does not occur in the field. 
What might it be? There are many possibilities: 1)
stars in the field were born in much shallower clusters which were rapidly destroyed when the first supernovae
appeared. This disruption process prevents a second generation of stars to appear; 2)
in small easily destroyed clusters, too few massive stars are present, thus the source of CNO-rich material is 
too weak for allowing many anomalous stars to be formed; 3) another possibility would be that the rotation rate of massive stars depends somewhat on the strength of the star formation event. This is of course quite speculative, but it is interesting to note that the rotation rate of B-type stars in clusters seems to be higher than that of B-type stars in the field \citep{HuangGies2006}.

Even if anomalous stars are formed exclusively in clusters, one can still wonder why some of them could not have escaped the cluster and passed the rest of their life in the field. It is reasonable to think that
during their lifetime the globular clusters lose part of their stellar content by ``evaporation'' processes
like for instance tidal disruption. Thus, why no anomalous stars are lost in that way? As explained in our scenario II, this may arise due to the fact that anomalous stars are preferentially formed in the central parts
of the cluster and are still in that region when the major evaporating events, affecting mostly the stellar population lying in the outer layers of the cluster, occur.

It might be also that anomalous stars are actually present in the field, but in such a small proportion that they have until now escaped detection. Let us estimate an upper limit to the fraction of anomalous stars in the field, $f_A$. If no anomalous
stars are detected among $N_{\rm obs}$ field stars,
it means that $f_A N_{\rm obs} << 1$ or that $f_A << 1/N_{\rm obs}$. 
In case $N_{\rm obs}\sim 100$, much less than one star over 
hundred in the field is anomalous.

From $N_{\rm obs}$ and the degree of evaporation (see below), we can
constrain $F_A$, the fraction of stars lost by globular 
clusters which are anomalous.
Let us recall that
the number of stars in the halo
is around 10$^{9}$. The number of stars in present-day globular cluster amounts to
about 1\% of the total number of stars in the halo \citep{Woltjer1975}. Let us say that
the globular clusters have lost 
$K$ times the number of stars they contain today.
This implies that the fraction of stars now in the field but having been born in globular clusters
is $f=K 0.01$. 
If $K=10$
({\it i.e.} if globular clusters lost about 10 times the number of stars they contain today), then one field star over 10
would have been born in a globular cluster.
Then the fraction of anomalous stars in the field, $f_A$, can be written
$f_A={F_A K \over 100} << 1/N_{\rm obs}$. This implies that
$F_A << {100 \over K N_{\rm obs}}$. With 
$K=10$, and $N_{\rm obs}\sim 100$, $F_A<<0.10$.
Increasing $N_{\rm obs}$  will provide tighter constraint on $F_A$ (provided $K$
is known from other considerations). From $N_{\rm obs}$,
an IMF, as well as an observed averaged fraction of anomalous to normal stars in globular clusters
some constrains on the degree of evaporation of first and second generation stars
can be obtained.
 
\subsection{Supernova rates in very young globular clusters}

In case the chemical anomalies are linked to a population of massive stars, what would be the consequences on the expected rate of supernovae in the early evolutionary phases of globular clusters?
With a slope of the IMF equal to  0.55 as in our scenario II, one would expect a supernova rate of 1 to 2 per 1000 year per globular cluster (containing initially one million stars with masses distributed between 0.01 and 120 M$_\odot$). This rate would be maintained for a few million years (about 8 millions years). If the halo of a galaxy contains about 100 globular clusters, then, one would expect 1 to 2 supernovae per 10 years observing the system composed by all globular clusters.
Interestingly, part of these supernovae, those originating from initial masses superior to about 50 M$_\odot$
would likely give birth to collapsars and then to Gamma Ray Bursts due their fast rotation. During about 1.5 million years, in the very early phases of the globular cluster (typically with ages between 3.5 and 5 million years) one expects a rate of GRB of about 2 per 1000 years per globular cluster, or 2 per 10 years when the whole system of globular clusters of a galaxy is surveyed.
In case the slope of the IMF is 1.35, the expected rates are lowered by about three orders of magnitudes. Typically the GRB rate would be between 0.002 and 0.004 per 1000 years and per globular clusters.

\subsection{He-rich stars and the UV-upturn in ellipticals}

Interestingly, all ellipticals show the UV-upturn phenomenon, i.e. an UV excess in their spectra. These excesses are caused by an old population of hot helium-burning stars.
Among the models which have been proposed to explain this UV-upturn, let us mention \citet{Lee1994,Parklee1997} who propose that the blue core-helium-burning stars, originate from the low-metallicity tail of a stellar population with a wide metallicity distribution. \citet{Bressanetal94, Yietal97} on the contrary, assumes that the UV-upturn is caused by metal-rich stars that lose their envelope near the tip of the first giant branch. More recently, \citet{Hanetal2007} propose that these stars lost their envelope because of binary interactions. 

As seen above, He-rich stars seem to be present in globular clusters (already on the Main Sequence).
As in elliptical galaxies, some globular clusters also present an extended blue horizontal branch.
In globular clusters this hot branch of helium-burning stars is 
believed to arise from the He-rich stellar population.
Could it be also the case in elliptical galaxies? Or said in other words, do ellipticals also contain
a population of non-evolved He-rich low mass stars, whose origin may be linked with a previous generations of 
fast rotating massive stars?

In the monolithic formation scenario, elliptical galaxies underwent at their birth a very strong burst of star formation.
If there is some link between strong star formation episodes and the presence of very fast rotators, then
one would expect in elliptical galaxies a great number of massive fast rotators. 
Would these fast rotators be able to produce anomalous stars in ellipticals?
A big difference between globular clusters and ellipticals is the depth of the gravitational potential well.
In ellipticals, the slow and fast winds and the supernovae ejecta are retained (at least in part). Thus the circumstances are less favorable to the apparition of stars, made of a mixture of interstellar medium and of only H-burning products\footnote{Note that provided only wind material is used (both slow and fast winds), very Helium rich stars can be formed. Inclusion of supernova ejecta however (without strong fallback) would prevent reaching values of 
$\Delta Y/\Delta Z$ as high 70, \citep[see Fig.~3 in][]{MMocen}}.
However this might not be correct and the argument is the following: in our scenario for globular clusters, we must assume that the star formation occurs rapidly in the vicinity of the massive stars when the slow winds is released. If not, the fast winds and the supernovae explosions would remove this interstellar material. Interestingly, clumps or protostars are observed in the disk of the (Herbig) Be star MWC 1080
\citep{Wang07}, indicating at least that low mass star formation can occur at a very early stage in an equatorial circumstellar disk.  
Thus in ellipticals, even if fast winds and supernova ejecta would be retained, provided the formation of
clumps in equatorial disk occur rapidly we still could have stars with strong helium enrichments and bearing
the signature of CNO-processed material.
Thus  a similar scenario than the one
suggested above for explaining the anticorrelation in globular clusters may have occurred in elliptical galaxies. This may have interesting consequences on the early phases of the photometric and chemical evolution of the massive elliptical galaxies.

%

\bibliographystyle{aa}   
\def\aj{AJ}%
\def\actaa{Acta Astron.}%
\def\araa{ARA\&A}%
\def\apj{ApJ}%
\def\apjl{ApJ}%
\def\apjs{ApJS}%
\def\ao{Appl.~Opt.}%
\def\apss{Ap\&SS}%
\def\aap{A\&A}%
\def\aapr{A\&A~Rev.}%
\def\aaps{A\&AS}%
\def\azh{AZh}%
\def\baas{BAAS}%
\def\bac{Bull.~astr.~Inst.~Czechosl.}%
\def\caa{Chinese Astron. Astrophys.}%
\def\cjaa{Chinese J. Astron. Astrophys.}%
\def\icarus{Icarus}%
\def\jcap{J. Cosmology Astropart. Phys.}%
\def\jrasc{JRASC}%
\def\mnras{MNRAS}%
\def\memras{MmRAS}%
\def\na{New A}%
\def\nar{New A Rev.}%
\def\pasa{PASA}%
\def\pra{Phys.~Rev.~A}%
\def\prb{Phys.~Rev.~B}%
\def\prc{Phys.~Rev.~C}%
\def\prd{Phys.~Rev.~D}%
\def\pre{Phys.~Rev.~E}%
\def\prl{Phys.~Rev.~Lett.}%
\def\pasp{PASP}%
\def\pasj{PASJ}%
\def\qjras{QJRAS}%
\def\rmxaa{Rev. Mexicana Astron. Astrofis.}%
\def\skytel{S\&T}%
\def\solphys{Sol.~Phys.}%
\def\sovast{Soviet~Ast.}%
\def\ssr{Space~Sci.~Rev.}%
\def\zap{ZAp}%
\def\nat{Nature}%
\def\iaucirc{IAU~Circ.}%
\def\aplett{Astrophys.~Lett.}%
\def\apspr{Astrophys.~Space~Phys.~Res.}%
\def\bain{Bull.~Astron.~Inst.~Netherlands}%
\def\fcp{Fund.~Cosmic~Phys.}%
\def\gca{Geochim.~Cosmochim.~Acta}%
\def\grl{Geophys.~Res.~Lett.}%
\def\jcp{J.~Chem.~Phys.}%
\def\jgr{J.~Geophys.~Res.}%
\def\jqsrt{J.~Quant.~Spec.~Radiat.~Transf.}%
\def\memsai{Mem.~Soc.~Astron.~Italiana}%
\def\nphysa{Nucl.~Phys.~A}%
\def\physrep{Phys.~Rep.}%
\def\physscr{Phys.~Scr}%
\def\planss{Planet.~Space~Sci.}%
\def\procspie{Proc.~SPIE}%

\bibliography{bibADS}

\begin{thebibliography}{28}
\expandafter\ifx\csname natexlab\endcsname\relax\def\natexlab#1{#1}\fi

\bibitem[{{Anderson}(1997)}]{Anderson97}
{Anderson}, A.~J. 1997, PhD thesis, AA(UNIVERSITY OF CALIFORNIA, BERKELEY)

\bibitem[{{Bonifacio} {et~al.}(2007){Bonifacio}, {Pasquini}, {Molaro},
  {Carretta}, {Fran{\c c}ois}, {Gratton}, {James}, {Sbordone}, {Spite}, \&
  {Zoccali}}]{BonifacioPasquini2007}
{Bonifacio}, P., {Pasquini}, L., {Molaro}, P., {et~al.} 2007, ArXiv e-prints,
  704

\bibitem[{{Bressan} {et~al.}(1994){Bressan}, {Chiosi}, \&
  {Fagotto}}]{Bressanetal94}
{Bressan}, A., {Chiosi}, C., \& {Fagotto}, F. 1994, \apjs, 94, 63

\bibitem[{{Caloi} \& {D'Antona}(2007)}]{Caloi07}
{Caloi}, V. \& {D'Antona}, F. 2007, \aap, 463, 949

\bibitem[{{Carretta}(2006)}]{Carretta2006}
{Carretta}, E. 2006, \aj, 131, 1766

\bibitem[{{Carretta} {et~al.}(2007){Carretta}, {Bragaglia}, {Gratton},
  {Lucatello}, \& {Momany}}]{CarrettaBragaglia2007}
{Carretta}, E., {Bragaglia}, A., {Gratton}, R., {Lucatello}, S., \& {Momany},
  Y. 2007, \aap{} accepted, ArXiv Astrophysics e-prints: astro-ph/0701174

\bibitem[{{Charbonnel}(2005)}]{Charbonnel2005}
{Charbonnel}, C. 2005, in IAU Symposium, ed. V.~{Hill}, P.~{Fran{\c c}ois}, \&
  F.~{Primas}, 347--356

\bibitem[{{Cohen} \& {Mel{\'e}ndez}(2005)}]{CohenMelendez2005}
{Cohen}, J.~G. \& {Mel{\'e}ndez}, J. 2005, \aj, 129, 1607

\bibitem[{{Decressin} {et~al.}(2007{\natexlab{a}}){Decressin}, {Charbonnel}, \&
  {Meynet}}]{DecressinII}
{Decressin}, T., {Charbonnel}, C., \& {Meynet}, G. 2007{\natexlab{a}}, \aap,
  475, 859

\bibitem[{{Decressin} {et~al.}(2007{\natexlab{b}}){Decressin}, {Meynet},
  {Charbonnel}, {Prantzos}, \& {Ekstr{\"o}m}}]{DecressinI}
{Decressin}, T., {Meynet}, G., {Charbonnel}, C., {Prantzos}, N., \&
  {Ekstr{\"o}m}, S. 2007{\natexlab{b}}, \aap, 464, 1029

\bibitem[{{Ekstr{\"o}m} {et~al.}(2007){Ekstr{\"o}m}, {Meynet}, {Maeder}, \&
  {Barblan}}]{Ekstrom07}
{Ekstr{\"o}m}, S., {Meynet}, G., {Maeder}, A., \& {Barblan}, F. 2007, ArXiv
  e-prints, 711

\bibitem[{{Gratton} {et~al.}(2004){Gratton}, {Sneden}, \&
  {Carretta}}]{GrattonSneden2004}
{Gratton}, R., {Sneden}, C., \& {Carretta}, E. 2004, \araa, 42, 385

\bibitem[{{Han} {et~al.}(2007){Han}, {Podsiadlowski}, \&
  {Lynas-Gray}}]{Hanetal2007}
{Han}, Z., {Podsiadlowski}, P., \& {Lynas-Gray}, A.~E. 2007, \mnras, 380, 1098

\bibitem[{{Huang} \& {Gies}(2006)}]{HuangGies2006}
{Huang}, W. \& {Gies}, D.~R. 2006, \apj, 648, 580

\bibitem[{{Kaviraj} {et~al.}(2007){Kaviraj}, {Sohn}, {O'Connell}, {Yoon},
  {Lee}, \& {Yi}}]{Kaviraj07}
{Kaviraj}, S., {Sohn}, S.~T., {O'Connell}, R.~W., {et~al.} 2007, \mnras, 377,
  987

\bibitem[{{Lee}(1994)}]{Lee1994}
{Lee}, Y.-W. 1994, \apjl, 430, L113

\bibitem[{{Maeder}(1992)}]{Maeder92}
{Maeder}, A. 1992, \aap, 264, 105

\bibitem[{{Maeder} \& {Meynet}(2006)}]{MMocen}
{Maeder}, A. \& {Meynet}, G. 2006, \aap, 448, L37

\bibitem[{{Norris}(2004)}]{Norris04}
{Norris}, J.~E. 2004, \apjl, 612, L25

\bibitem[{{Pagel} {et~al.}(1992){Pagel}, {Simonson}, {Terlevich}, \&
  {Edmunds}}]{Pagel92}
{Pagel}, B.~E.~J., {Simonson}, E.~A., {Terlevich}, R.~J., \& {Edmunds}, M.~G.
  1992, \mnras, 255, 325

\bibitem[{{Park} \& {Lee}(1997)}]{Parklee1997}
{Park}, J.-H. \& {Lee}, Y.-W. 1997, \apj, 476, 28

\bibitem[{{Pasquini} {et~al.}(2005){Pasquini}, {Bonifacio}, {Molaro},
  {Francois}, {Spite}, {Gratton}, {Carretta}, \&
  {Wolff}}]{PasquiniBonifacio2005}
{Pasquini}, L., {Bonifacio}, P., {Molaro}, P., {et~al.} 2005, \aap, 441, 549

\bibitem[{{Piotto} {et~al.}(2005){Piotto}, {Villanova}, {Bedin}, {Gratton},
  {Cassisi}, {Momany}, {Recio-Blanco}, {Lucatello}, {Anderson}, {King},
  {Pietrinferni}, \& {Carraro}}]{Piotto05}
{Piotto}, G., {Villanova}, S., {Bedin}, L.~R., {et~al.} 2005, \apj, 621, 777

\bibitem[{{Porter} \& {Rivinius}(2003)}]{PorterRivinius2003}
{Porter}, J.~M. \& {Rivinius}, T. 2003, \pasp, 115, 1153

\bibitem[{{Prantzos} \& {Charbonnel}(2006)}]{PC06}
{Prantzos}, N. \& {Charbonnel}, C. 2006, \aap, 458, 135

\bibitem[{{Wang} {et~al.}(2007){Wang}, {Looney}, {Brandner}, \&
  {Close}}]{Wang07}
{Wang}, S., {Looney}, L.~W., {Brandner}, W., \& {Close}, L.~M. 2007, ArXiv
  e-prints, 709

\bibitem[{{Woltjer}(1975)}]{Woltjer1975}
{Woltjer}, L. 1975, \aap, 42, 109

\bibitem[{{Yi} {et~al.}(1997){Yi}, {Demarque}, \& {Kim}}]{Yietal97}
{Yi}, S., {Demarque}, P., \& {Kim}, Y.-C. 1997, \apj, 482, 677

\end{thebibliography}


\end{document}